\documentclass[12pt,preprint]{aastex}

\begin{document}

\title{3D Hydrodynamic Simulations of Relativistic Extragalactic Jets}
\author{Philip A. Hughes}
\affil{Astronomy Department, University of Michigan, Ann Arbor, MI 48109-1090}
\email{hughes@astro.lsa.umich.edu}
\author{Mark A. Miller}
\affil{Department of Physics, Washington University, St. Louis, MO  63130-4899}
\email{ mamiller@wugrav.wustl.edu}
\and
\author{G. Comer Duncan}
\affil{Department of Physics and Astronomy, Bowling Green State University, Bowling Green, OH 43403}
\email{ gcd@chandra.bgsu.edu}

\begin{abstract}
 We describe a new numerical 3D relativistic hydrodynamical code and present
the results of three validation tests.  A comparison of an axisymmetric jet
simulation using the 3D code, with corresponding results from an earlier 2D
code, reveal that a) the enforcement of axisymmetry in the 2D case had no
significant influence on the global morphology and dynamics; b) although 3D
studies typically have lower resolution than those using 2D, limiting their
ability to fully capture internal jet structure, such 3D studies can provide
a reliable model of global morphology and dynamics.

 The 3D code has been used to study the deflection and precession of
relativistic flows. We find that even quite fast jets ($\gamma\sim 10$)
can be significantly influenced by impinging on an oblique density gradient,
exhibiting a rotation of the Mach disk in the jet's head. The flow is bent
via a potentially strong, oblique internal shock that arises due to
asymmetric perturbation of the flow by its cocoon. In extreme cases this
cocoon can form a marginally relativistic flow {\it orthogonal} to the jet,
leading to large scale dynamics quite unlike that normally associated with
astrophysical jets. Exploration of a $\gamma=5$ flow subject to a large
amplitude precession (semi-angle $11\fdg 25$) shows that it retains its
integrity, with modest reduction in Lorentz factor and momentum flux, for
almost $50$ jet-radii, but thereafter, the collimated flow is disrupted.  The
flow is approximately ballistic, with velocity vectors {\it not} aligned with
the local jet `wall'. However, sufficiently large changes in flow direction
take place within the jet that for observers close to the jet axis,
significant changes in Doppler boost would be evident along the flow.

 We consider simple estimators of the flow emissivity in each case and
conclude that a) while the oblique internal shocks which mediate a small
change in the direction of the deflected flows have little impact on the
global dynamics, significantly enhanced flow emission (by a factor of $2-3$)
may be associated with such regions; and b) the convolution of rest frame
emissivity and Doppler boost in the case of the precessed jet invariably
leads to a core-jet-like structure, but that intensity fluctuations in the
jet cannot be uniquely associated with either change in internal conditions
or Doppler boost alone, but in general are a combination of both factors.
\end{abstract}


\section{Introduction}

Collimated extragalactic flows, exhibiting a complex pattern of internal
structures, often with high brightness temperature and/or superluminal speed,
have been explored with increasing spatial and temporal resolution over the
last two decades \citep{zen}.  Remarkably, it has become evident over the
last five years that the highly energetic flows associated with active
galactic nuclei are also found in Galactic objects with stellar mass
`engines'. In the galactic superluminals GRS~1915+105 \citep{mr1,mr2} and
GRO~J1655--40 \citep{hje,tin} the observed motions indicate jet flow speeds
up to $0.98c$.  Further, there is compelling evidence from the observation of
optical afterglow that gamma-ray bursts are of cosmological origin, and
whether produced by the mergers of compact objects, or accretion induced
collapse (AIC), simple relativistic fireball models seem ruled out, the data
strongly favoring highly relativistic jets \citep{dar}.  Thus a detailed
understanding of the dynamics of collimated relativistic flows has wide
application in astrophysics.

A particularly important facet of these flows has been revealed over the last
decade as the quantity and quality of images of extragalactic relativistic
jets have increased: the curvature of parsec-scale flows is the norm, not the
exception; see  e.g., \citet{war}.  Some of this curvature may be `apparent',
resulting from a more modestly curved flow seen close to the line of sight.
Nevertheless, such flows must posses intrinsic curvature, and we are forced
to ask how such highly relativistic flows can exhibit significant curvature,
and yet retain their integrity, transporting energy and momentum efficiently
to the kiloparsec scale and beyond.  Indeed, the structure (large-scale
curvature, sheaths and filaments, and `knots' of emission) of such flows
provides a powerful diagnostic tool, that may elucidate the intrinsic
instability, precession, deflecting obstacles and cross-winds to which the
flow is subjected (although it is important to note that the mere presence
of structure cannot uniquely determine its origin: at a minimum we must
explore the velocity and magnetic fields of such features, and their
evolution).

Numerous lines of evidence point convincingly to the occurrence of both
transverse and oblique shocks, e.g., \citep{hab,lam,maa,paw} in such flows;
how do they form and evolve? Specifically, how do transverse shocks propagate
along a curved flow; will the shock plane rotate with respect to the flow
axis; will the shock strengthen or weaken?  What role does flow curvature
have to play in explaining phenomena such as stationary knots between which
superluminal components propagate \citep{sha}, knots which brighten after an
initial fading \citep{mut}, and changing component speed \citep{laz}.  All
these issues are potentially important in both the extragalactic and stellar
jet contexts, and are amenable to study through hydrodynamic simulation --
but all demand that such simulations be 3D.

Numerical hydrodynamical studies of relativistic astrophysical jets have been
available for barely five years. \citet{wil} and \citet{bow} explored {\it
steady} relativistic jets, but it was only with the advent of robust,
shock-capturing 2D schemes \citep{vp1,vp2,dah,ma1,ma2,ma3,kf2,ros} that a
full exploration of relativistic flows began.  \citet{koi}, \citet{ni1} and
\citet{ni2} have pioneered 3D MHD studies where fluid is injected into a
strong, ordered magnetic field, but the detailed structure and evolution of
3D flows remains largely unexplored \citep{alo}. The current paper describes
extension of our original 2D code \citep{dah} to 3D, and its first
application to axisymmetric, deflected and precessed flows. Our immediate
goals are to study: the internal structures that arise from the development
of Kelvin-Helmholtz instability, which may explain features such as those
seen in the jet of M~87 (e.g., components E and A \citep{per}); the formation
of oblique shocks which may mediate a change in flow direction (e.g.,
1127$-$-145 and CTA~102 \citep{jor}); and precession which may underly the
{\it evolving}, approximately helical trajectories exhibited by a number of
parsec-scale flows (e.g., BL~Lac \citep{dmm}).

\section{Numerical Solver for the Euler Equations}

 We assume an inviscid and compressible gas, and an ideal equation of state
with constant adiabatic index. We use a Godunov-type solver which is a
relativistic generalization of the method due to \citet{hlv}, and
\citet{ein}, in which the full solution to the Riemann problem is
approximated by two waves separated by a piecewise constant state.  We evolve
mass density $R$, the three components of the momentum density $M_x$, $M_y$
and $M_z$, and the total energy density $E$ relative to the laboratory
frame.

Defining the vector (in terms of its transpose for compactness)
\begin{equation}
U = (R,M_x,M_y,M_z,E)^{T} ,
\end{equation}
and the three flux vectors
\begin{equation}
F^x = (R v^x,M_x v^x + p, M_y v^y, M_z v^z, (E + p)v^x) ^{T } ,
\end{equation}
\begin{equation}
F^y = (R v^y,M_x v^x, M_y v^y + p, M_z v^z, (E + p)v^y) ^{T } ,
\end{equation}
\begin{equation}
F^z = (R v^z,M_x v^x, M_y v^y, M_z v^z + p,(E + p)v^z) ^{T } ,
\end{equation}
the conservative form of the relativistic Euler equation is
\begin{equation}
{\partial{ U}\over \partial{t}} + {\partial\over \partial{x} } (F^x) +
{\partial\over \partial y} (F^{y})+ {\partial\over \partial z} (F^{z}) = 0 .
\end{equation}
The pressure is given by the ideal gas equation of state $ p = (\Gamma - 1)
(e - n) . $ The Godunov-type solvers are well known
for their capability as robust, conservative flow solvers with excellent
shock capturing features.  In this family of solvers one reduces the problem
of updating the components of the vector $U$, averaged over a cell, to the
computation of fluxes at the cell interfaces. In one spatial dimension the
part of the update due to advection of the vector $U$ may be written as
\begin{equation}
{U^{n+1}}_{i} = {U^{n}}_{i} - \frac{\delta t}{\delta x} (F_{i + \frac{1}{2} } -
F_{i - \frac{1}{2}}) .
\end{equation}
In the scheme originally devised by \citet{god}, a fundamental emphasis is
placed on the strategy of decomposing the problem into many local Riemann
problems, one for each pair of values of $U_{i}$ and $U_{i+1}$, to yield
values which allow the computation of the local interface fluxes
$F_{i+\frac{1}{2}}$.  In general, an initial discontinuity at $i+\frac{1}{2}$
due to $U_{i}$ and $U_{i+1}$ will evolve into four piecewise constant states
separated by three waves. The left-most and right-most waves may be either
shocks or rarefaction waves, while the middle wave is always a contact
discontinuity. The determination of these four piecewise constant states can,
in general, be achieved only by iteratively solving nonlinear equations.
Thus the computation of the fluxes necessitates a step which can be
computationally expensive.  For this reason much attention has been given to
approximate, but sufficiently accurate, techniques.  One notable method is
that due to \citet[HLL]{hlv}, in which the middle wave, and the two constant
states that it separates, are replaced by a single piecewise constant state.
One benefit of this approximation, which smears the contact discontinuity
somewhat, is to eliminate the iterative step, thus significantly improving
efficiency.  However, the HLL method requires accurate estimates of the wave
speeds for the left- and right-moving waves.  \citet{ein} analyzed the HLL
method and found good estimates for the wave speeds. The resulting method
combining the original HLL method with Einfeldt's improvements (the HLLE
method),  has been taken as a starting point for our simulations.  In our
implementation we use wave speed estimates based on a simple application of
the relativistic addition of velocities formula for the individual components
of the velocities, and the relativistic sound speed $c_s$, assuming that the
waves can be decomposed into components moving perpendicular to the three
coordinate directions.

In order to compute the pressure $p$ and sound speed $c_s$ we need the rest
frame mass density $n$ and energy density $e$.  However, these quantities are
nonlinearly coupled to the components of the velocity as well as to the
laboratory frame variables via the Lorentz transformation:
\begin{equation}
R = \gamma n ,
\end{equation}
\begin{equation}
M^{x} = \gamma^2 ( e + p ) v^{x} ,
\end{equation}
\begin{equation}
M^{y} = \gamma^2 ( e + p ) v^{y} ,
\end{equation}
\begin{equation}
M^{z} = \gamma^2 ( e + p ) v^{z} ,
\end{equation}
\begin{equation}
E = \gamma^2 ( e + p ) - p ,
\end{equation}
where $\gamma = ( 1 - v^2 )^{-1/2}$ is the Lorentz factor and $v^2 =
(v^{x})^2 + (v^{y})^2 + (v^{z})^2$. When the adiabatic index is constant it
is possible to reduce the computation of $n$, $e$, $v^{x}$, $v^{y}$ and
$v^{z}$ to the solution of the following quartic equation:
\begin{eqnarray}
& & \Bigl\lbrack\Gamma v \left(E-Mv\right)-M  \left(1-v^2\right)\Bigr\rbrack^2
 \nonumber \\
& & -  \left(1-v^2\right)v^2\left(\Gamma-1\right)^2R^2=0 ,
\end{eqnarray}
where $M^2 = (M^{x})^2 + (M^{y})^2 + (M^{z})^2$. This quartic is solved at
each cell several times during the update of a given mesh using
Newton-Raphson iteration.

Our scheme is generally of second order accuracy, which is achieved by taking
the state variables as piecewise linear in each cell, and computing fluxes at
the half-time step. However, in estimating the laboratory frame values on
each cell boundary, it is possible that through discretization, the
laboratory frame quantities are unphysical -- they correspond to rest frame
values $v>1$ or $p<0$. At each point where a transformation is needed, we
check that the conditions $M/E<1$ and
$R/E<\left[1-\left(M/E\right)^2\right]^{1/2}$ are satisfied, and if not,
recompute cell interface values in the piecewise constant approximation.  We
find that such `fall back to first order' rarely occurs.

\section{Adaptive Mesh Refinement}

The relativistic HLLE (RHLLE) method constitutes the basic flow integration
scheme on a single mesh.  We use adaptive mesh refinement (AMR) in order to
gain spatial and temporal resolution for given computer resources.

The AMR algorithm used is a general purpose mesh refinement scheme which is
an outgrowth of original work by \citet{ber} and \citet{bac}. The AMR method
uses a hierarchical collection of grids consisting of embedded meshes to
discretize the flow domain.  We have used a scheme which subdivides the
domain into logically rectangular meshes with uniform spacing in the three
coordinate directions, and a fixed refinement ratio of $\times 3$. The AMR
algorithm orchestrates i) the flagging of cells which need further
refinement, assembling collections of such cells into meshes; ii) the
construction of boundary zones so that a given mesh is a self-contained
entity consisting of the interior cells and the needed boundary information;
iii) mechanisms for sweeping over all levels of refinement and over each mesh
in a given level to update the physical variables on each such mesh; and iv)
the transfer of data between various meshes in the hierarchy, with the
eventual completed update of all variables on all meshes to the same final
time level. The adaption process is dynamic so that the AMR algorithm places
further resolution where and when it is needed, as well as removing
resolution when it is no longer required.  

Adaption occurs in time, as well as in space: the time step on a refined grid
is less than that on the coarser grid, by the refinement factor for the
spatial dimension. More time steps are taken on finer grids, and the advance
of the flow solution is synchronized by interleaving the integrations at
different levels. This helps prevent any interlevel mismatches that could
adversely affect the accuracy of the simulation. The time step value is
computed by applying the CFL condition to every cell within the computational
domain, using the relativistic sum of the local flow velocity and sound speed,
and picking the globally minimum value of $dt$. The corresponding time step
on the coarsest level is computed, and multiplied by a CFL number, typically
$0.1$.

In order for the AMR method to sense where further refinement is needed, some
monitoring function is required.  We have used a combination of the gradient
of the laboratory frame mass density, a test that recognizes the presence of
contact surfaces, and a measure of the cell-to-cell shear in the flow, the
choice of which functions to use being determined by which part of the flow
is of most significance in a given study. Since the tracking of shock waves
is of paramount importance, a buffer ensures the flagging of extra cells at
the edge of meshes, ensuring that important flow structures do not `leak' out
of refined meshes during the update of the hydrodynamic solution. The
combined effect of using the RHLLE single mesh solver and the AMR algorithm
results in a very efficient scheme.  Where the RHLLE method is unable to give
adequate resolution on a single coarse mesh the AMR algorithm places more
cells, resulting in an excellent overall coverage of the computational
domain.

\section{Code Validation}

 The testing of codes for both 2D and 3D numerical hydrodynamics, and the
appraisal of the quality of the `solver', in terms of its ability to
accurately capture both the structure and location of shocks and contact
surfaces is always challenging, because of the sparsity of problems with
analytic solutions; this is particularly true in the case of relativistic
flows. We present the results of three common tests below, but also note that
the solver employed in the current code is a direct extension to 3D (with a
recast from Fortran 77 to Fortran 90) of the solver described by
\citet{dah}.  Evidence for the accuracy and robustness of that code comes
from, in addition to its application to test problems: a) the general
agreement between studies performed with that code and with independently
constructed codes, e.g., that of \citet{ma3}; b) the agreement between
simulations performed with that code and analytic estimates of the expected
morphology and dynamics, e.g., \citet{ros}; and c) the agreement between
internal jet structures found in simulations performed with that code and the
predictions of linear stability analyses \citep{har,ros}.  The primary goal
of the tests presented here is to validate the extension and translation of
the original 2D code, and its inclusion in a newly constructed AMR
`harness'.

\subsection{1D Relativistic Shock Tube}

 The initial condition for the 1D relativistic shock tube problem comprises
two piecewise constant states separated by a discontinuity at the origin.
The left and right states have pressure and rest density: $p_L=10^4$,
$p_R=10.0$, $n_L=1.0$, and $n_R=0.1$. The fluid is initially at rest and the
adiabatic index is taken to be $\Gamma=4/3$. For time $t>0$ a rarefaction
wave propagates to the left and a contact surface and shock wave, which bound
a narrow region of shocked flow, propagate to the right; the analytic
solution has been provided by \citet{tom}.

\begin{figure}
\epsscale{0.75}
\plotone{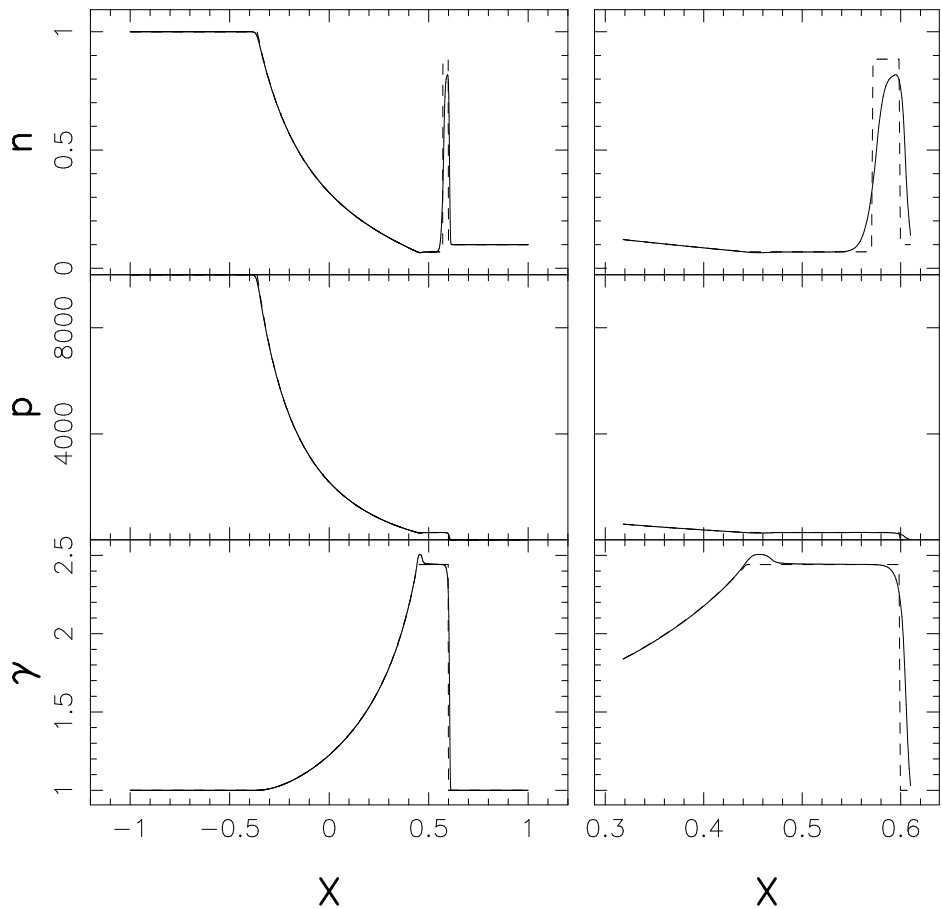}
\caption{
The relativistic shock tube.  The left column shows the entire computational
domain and the right panels show blow ups of the region encompassing the
shock. The solid line is the numerical solution for a shock propagating along
$x$, and the dashed line is the analytic solution from Thompson (1986). The
computation was taken to a time of 0.626 on a grid of 1365 cells. The top
panel shows the rest density ($n$), the middle panel the pressure ($p$), and
the bottom panel the Lorentz factor ($\gamma$).
\label{rst}}
\end{figure}

 The computational domain is taken to be $x,y,z = [-1.0,1.0]$.  We have run
the shock tube problem with the shock propagating along the $x$, $y$ and $z$
axes, in each case in `unigrid' mode (a single level, high resolution grid)
and with refinement by two levels of $\times 3$ each, with the base grid
resolution chosen so that the resolution of the finest grid matches that of
the unigrid runs.  The code produces comparable results independent of the
coordinate direction of the shock propagation, and -- within refined regions
-- an adaptive solution that is as good in quality of fit to the analytic
solution as is the unigrid case. (In general the unigrid and refined
solutions are not identical to machine accuracy, because it is possible that
in the refined case -- depending on the choice of parameters that control
where refinement occurs -- there are sections of the computational domain
that remain unrefined; such domains can lead to very small changes in the
location of refined structures.) Figure~\ref{rst} shows an example of the
numerical and analytic solutions for the 1D shock tube tests. The left panels
show, from top to bottom, the rest frame density, pressure and Lorentz factor
for the whole computational domain; the right panels show in detail the
structure of the flow in the vicinity of the contact/shock. The dashed line
is the analytic solution.

\placefigure{rst}

\subsection{3D Relativistic Shock Reflection}

The initial condition for the 3D relativistic shock reflection problem
comprises two piecewise constant states separated by a discontinuity at an
arbitrary radius of 0.5. The inner and outer states have pressure and rest
density:  $p_I=1.0$, $p_O=4.0$, $n_I=1.0$, and $n_O=4.0$. The fluid is
initially at rest and the adiabatic index is taken to be $\Gamma=4/3$.  The
values were chosen to permit comparison with the test run results of
\citet{wen} for a 1D shock-patching code.  For time $t>0$ a shock propagates
towards the origin, reflects, and moves outwards to interact with the contact
surface between the initial states.

\begin{figure}
\plotone{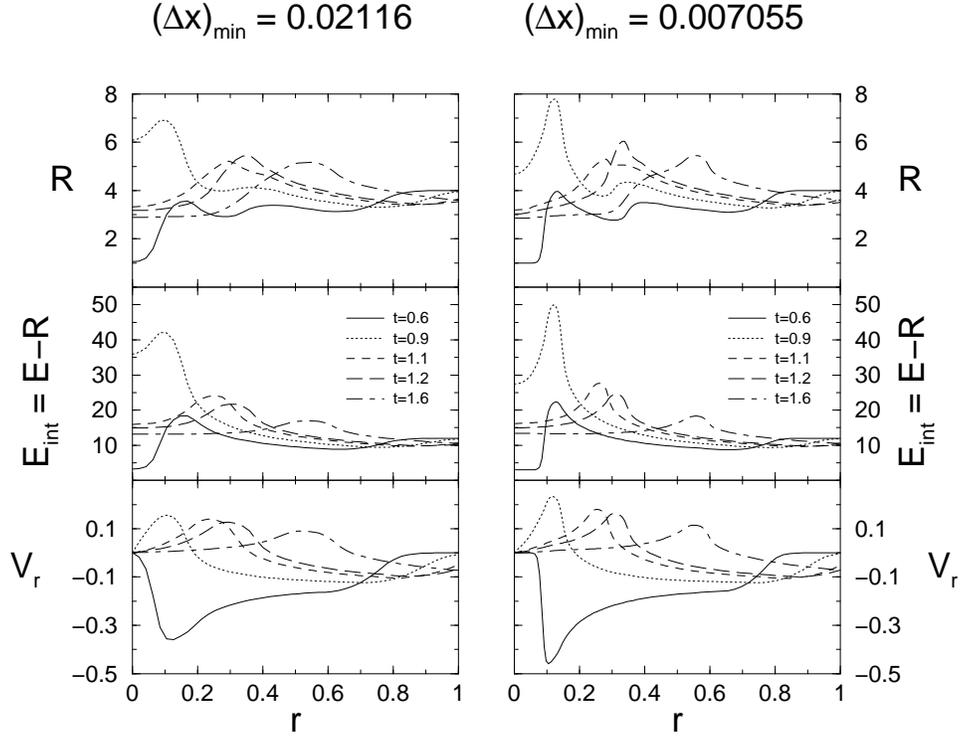}
\caption{
Cuts showing the evolution of the laboratory frame density ($R$), internal
energy ($E - R = {\gamma}^2 (e+p) - p - \gamma n$) and radial velocity ($v_r$)
for the relativistic shock reflection problem. In the left panels four levels
of refinement are used, in the right five levels are used. Within each
panel each line corresponds to a different elapsed time, as indicated by the
line-type code in the center panels.
\label{sph_refl}}
\end{figure}

The computational domain is taken to be $x = [-4.0,4.0]$, $y = [-4.0,4.0]$,
and $z = [-4.0,4.0]$.  The base grid (AMR level $= 1$) was chosen to contain
15 points, yielding a base grid discretization scale of ${\Delta
x}_{\verb+base+} = \frac{8}{14} \simeq 0.571$.  Note that the discretization
scale of the base grid is larger than the radius of the initial
discontinuity!  Figure~\ref{sph_refl} shows the evolution of the laboratory
frame density $R = \gamma n$, the laboratory frame internal energy density
$E_{\verb+int+} = E - R = {\gamma}^2 (e+p) - p - \gamma n$, and the
$x-$component of the fluid velocity $V^x$ along the $x-$axis.  The left panel
shows the evolution of the fields with three additional levels of refinement
(${\Delta x}_{\verb+fine+} = 8/14/3^3 \simeq 0.02116$), whereas the right
hand panel shows the evolution of the fluid with four additional levels of
refinement (${\Delta x}_{\verb+fine+} = 8/14/3^4 \simeq 0.007055$).  The
output times, $t=0.6, 0.9, 1.1, 1.2$, and $1.6$, were chosen so as to easily
compare with results from the $1-$D code written by Wen, Panaitescu, and
Laguna (Figure 4, right panel of ~\citet{wen}).  Although the discretization
scale for the $1-$D code ($\Delta r = 10^{-3}$) is almost an order of
magnitude smaller than the discretization scale of the finest grid of our 3D
AMR code (${\Delta x}_{\verb+fine+} = 0.007055$), we find good agreement
between the two.  Note that the effective resolution of this simulation is
1135~x~1135~x~1135, which would require over 200 GB (GigaBytes) of memory.
In contrast, our AMR code requires only 1.9 GB to run this simulation.  In
this case, AMR techniques have decreased the memory footprint of the
simulation by two orders of magnitude!

\placefigure{sph_refl}

\subsection{3D Relativistic Blastwave}

The 3D relativistic shock reflection problem does not exhibit highly
relativistic flow speeds during the evolution covered by the simulation, and
the major structures are interior to the location of the initial
discontinuity at $r=0.5$. In order to validate the code for flows $\gamma>>1$
in 3D and to explore the symmetry preserving properties of the code, we have
performed a 3D blastwave simulation. In this case a shell of high density and
pressure propagates outward from the initial discontinuity. Coding errors and
discretization effects rapidly become apparent as an asymmetric flow.
Ideally we would have used the Blandford-McKee blastwave solution \citep{bmk}
as initial data, confirming that while the shock speed remained highly
relativistic the computed solution matched the analytic one. However, when
using only a wedge-shaped portion of the entire sphere, of opening angle
$0.2$rad, and initial energy and density that yield a shock Lorentz factor
$\Gamma\sim 28.4$, and peak fluid Lorentz factor $\gamma=1/\sqrt{2}
\Gamma\sim 20$, the leading structure is so narrow that even five levels of
refinement (requiring $\sim 2$Gb of memory) did not adequately resolve it:
the highest Lorentz factor in the initial data being $\sim 8$. It was
concluded that the Blandford-McKee blastwave constitutes a problem that is
too demanding to be a useful validation test unless very significant computer
resources are employed.

We have thus opted to simulate the 3D-equivalent of the shock tube problem:
the initial condition for the 3D relativistic blastwave problem comprises two
piecewise constant states separated by a discontinuity at an arbitrary radius
$r=1.0$.  A numerical solution with spherical symmetry, and initial
conditions that produced marginally relativistic flow speed, was presented by
\citet{vp3}; here we adopt an initial state the leads to higher Lorentz
factors -- of order those encountered in the applications discussed in the
next sections. The inner and outer states have pressure and rest density:
$p_I=10^4$, $p_O=10.0$, $n_I=1.0$, and $n_O=0.1$. The fluid is initially at
rest and the adiabatic index is taken to be $\Gamma=4/3$. For time $t>0$ a
spherical shock wave propagates to larger radius. Substantial Lorentz factors
($\sim 5$) are generated, and by the end of the simulation the propagating
structure encompasses $\sim 35$ times the volume of the initial high pressure
sphere, which is ample development for isolating asymmetries that might
arise. Some asymmetry is to be expected, because the criteria used for
flagging cells to be refined is applied automatically, and a combination of
differences due to rounding, coupled with the clustering algorithm, leads to
a non-uniform distribution of refined patches.

\begin{figure}
\plotone{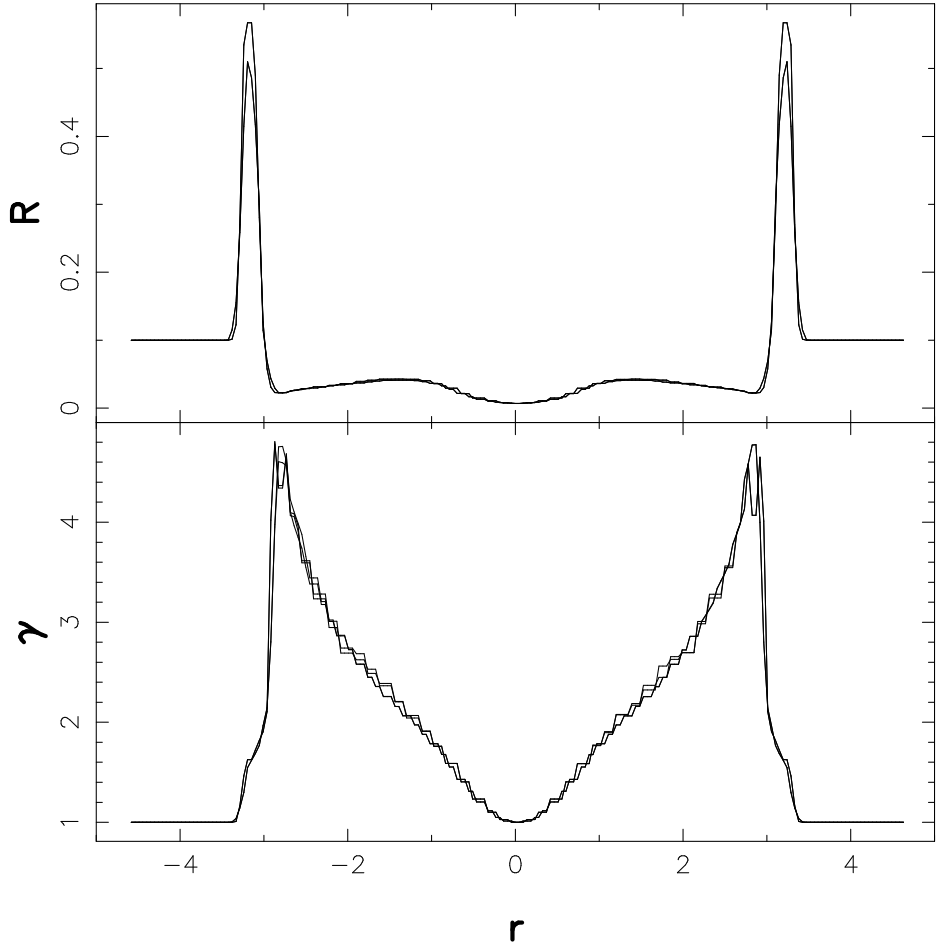}
\caption{
Cuts showing the laboratory frame density ($R$) and Lorentz factor ($\gamma$)
for the final time slice of the blastwave problem for cuts along the
coordinate directions and lines that bisect these directions.
\label{blast}}
\end{figure}

Taking the computational domain to be $x = [-10.0,10.0]$, $y = [-10.0,10.0]$,
and $z = [-10.0,10.0]$, Figure~\ref{blast} shows the final evolution (at
$t=2.3967$) of a run with $3$ levels of refinement, yielding $\sim 180$
finest level cells across the blastwave diameter at the end of the
computation. The two panels show laboratory frame density ($R$) and Lorentz
factor ($\gamma$) for cuts along the two orthogonal coordinate directions in
the plane $x=0$, plus additional cuts that bisect these two axes.  The peak
laboratory frame density along the diagonal cuts is $13\%$ higher than along
the coordinate directions.  This is associated with an oscillation in the
Lorentz factor near to its peak value $\sim 5$, in the cuts that bisect the
coordinate directions.  The oscillation in the Lorentz factor cuts is a
result of the fact that even with an effective resolution of $\sim 5.8 \times
10^6$ cells encompassing the blastwave, the scale of the leading edge of the
evolving structure is so narrow that the Lorentz factor varies from $\sim 5$
to $\sim 4$ over a scale of a couple of the finest cells.  Off axis, the
chance, and changing, location of cells with respect to the thin shell of
high Lorentz factor leads to a small amplitude irregularity in the peak
value, evident as fluctuation along the radial cut.  Notice however that an
inability to fully capture the extreme values encountered on the finest
scale, does not in general influence the global location of major flow
structures or their values away from these extremes. In particular, there is
no global asymmetry.

\placefigure{blast}

\subsection{Axisymmetric Jet Inflow}

It is well-known from nonrelativistic hydrodynamic studies that the most
rapidly growing modes of the Kelvin-Helmholtz instability are suppressed in
2D simulations (cf. \citet{hac}); only a fully 3D treatment can capture
features likely to play a major role in determining internal jet structure.
However, as internal flow structures thermalize only a small fraction of the
flow's bulk energy, axisymmetric simulations should provide a reliable
picture of the gross morphology and dynamics of an initially axisymmetric
flow. In order to test this expectation, and to learn what internal jet
structures arise in the absence of externally driven perturbations in 3D, we
have rerun case B of \citet{dah}, using the 3D code. If the 2D and 3D results
are similar in terms of gross morphology and dynamics, we can conclude that
these attributes are insensitive to the details of small scale flow
structure, and thus that a) earlier 2D results provided a valid picture of
these relativistic flows on the large scale; and b) relatively low resolution
3D simulations should provide a valid picture of the overall morphology and
dynamics of deflected and precessed jets.

The parameters for this simulation are (Lorentz factor) $\gamma=5.0$,
(relativistic Mach number) ${\cal M}=8$, and (adiabatic index) $\Gamma=5/3$.
The ambient density is $\times 10$ that of the inflowing jet rest density,
and the jet and ambient medium are initially in pressure balance. We have
achieved the same resolution within the jet ($24$ fine cells across the jet
radius) by using $3$ levels of refinement, of $\times 3$ each. However, to
avoid prohibitive memory requirements we have followed the jet for only $25$
jet-radii, as compared with $42$ jet-radii in the original 2D run, and in the
3D simulation we have used the finest grid only within $1.5$ jet-radii of the
axis -- thus the cocoon and bow are under-resolved.  The computation was
terminated as the bow shock reached the far edge of the domain.

\begin{figure}
\plotone{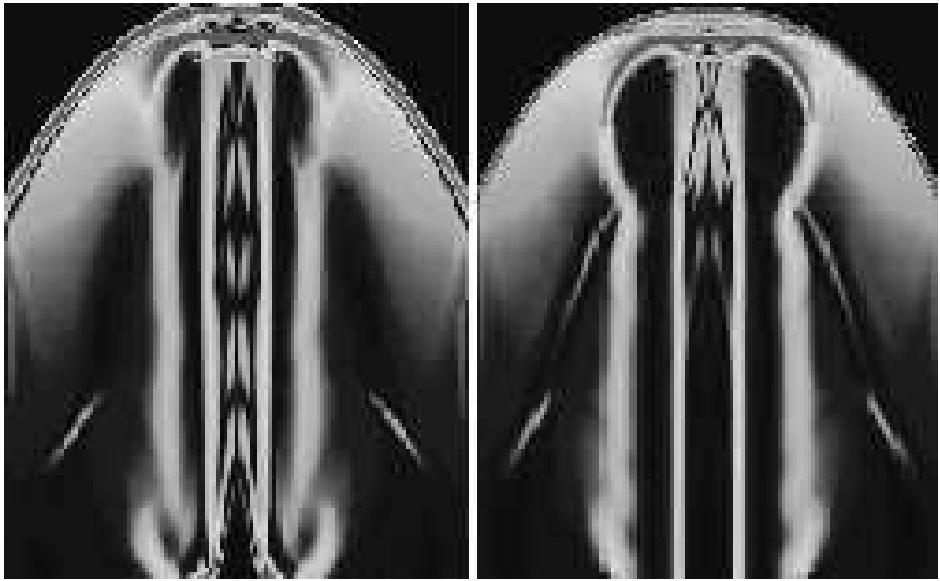}
\caption{
A schlieren plot of laboratory frame density gradient for an axisymmetric
inflow with $\gamma=5.0$, ${\cal M}=8$ and $\Gamma=5/3$ after 400 cycles.
Left: 3D simulation; Right: 2D simulation with the same resolution. As
discussed in the text, the slight differences in wave-structure between the
two simulations can be understood as due to a difference in the way the wave
modes are driven, in terms of the location and amplitude of the driving
perturbation, and its coupling.
\label{axicomp}}
\end{figure}

Figure~\ref{axicomp} shows a schlieren (gradient) rendition of the
laboratory frame density after 400 computational cycles.  For comparison, we
show also a 2D simulation performed with the same maximum resolution as that
achieved in 3D, having propagated $25$ jet-radii. In the absence of an
externally applied periodic perturbation to the flow, the development of
structure within the jet depends wholly on the driving of available modes by
naturally occurring perturbations -- in particular, pressure perturbations
arising from Kelvin-Helmholtz `fingers' growing at the contact surface
between shocked jet and shocked ambient medium, and a pressure wave at the
inflow \citep{har}.  In the original 2D simulation the contact surface
exhibited instability only as the bow approached the edge of the
computational domain. The poor resolution with which the contact is captured
in 3D, and the shorter duration of the simulation,  ensure that no
manifestation of this slowly growing instability is seen, and thus that
little jet structure develops as a result. However, the 3D simulation does
exhibit a much stronger pressure wave at inflow than is seen in the
corresponding 2D case. This is due to the radically different boundary
conditions necessary in 3D: inflow is imposed on the plane $z=$constant, and
involves cells cut by the jet boundary for which state variables must be
established through a volume-weighted average of the internal and external
values. To avoid a `leakage' of jet momentum into the ambient material,
fixed, initial values are used across the entire boundary plane at every time
step. In the 2D simulation -- above the jet inflow -- conditions were chosen
to be `extrapolated' or `rigorous outflow', but the initial conditions were
not enforced for the duration of the computation.  The ambient flow thus
evolved to smear out the shear layer near the inflow in the 2D case, reducing
the shear-induced wave in the jet. 

Thus for 2D and 3D simulations {\it of equal resolution} there is stronger
driving of available modes by a pressure wave at the inflow \citep{har} and
over most of its length the 3D simulation exhibits more internal structure;
the 2D simulation manifests internal structure of significant amplitude only
just upstream of the Mach disk, where the development of instability in the
better-resolved contact surface drives a weak pressure wave into the jet.
From a measurement of the well-defined locations of pressure minima along the
flow axis we estimate a wavelength for internal structure in the 3D case of
$\lambda^{obs}\sim 7$ jet radii.  As noted in \citet{har}, the actual jet
radius is somewhat larger than the inflow radius, and thus analytic
predictions based on the inflow scale slightly underestimate the wavelength
of excited modes. Bearing this in mind, it is plausible to identify the mode
excited in the 3D simulation as the second body mode, with
$\lambda^{max}_2\sim 4.6$. This is consistent with the fact that the lower
order modes have longer wavelength, and are oblique to the jet axis, and the
oblique inlet pressure wave more easily couples to these than to higher
order ones.

\placefigure{axicomp}

\begin{figure}
\plotone{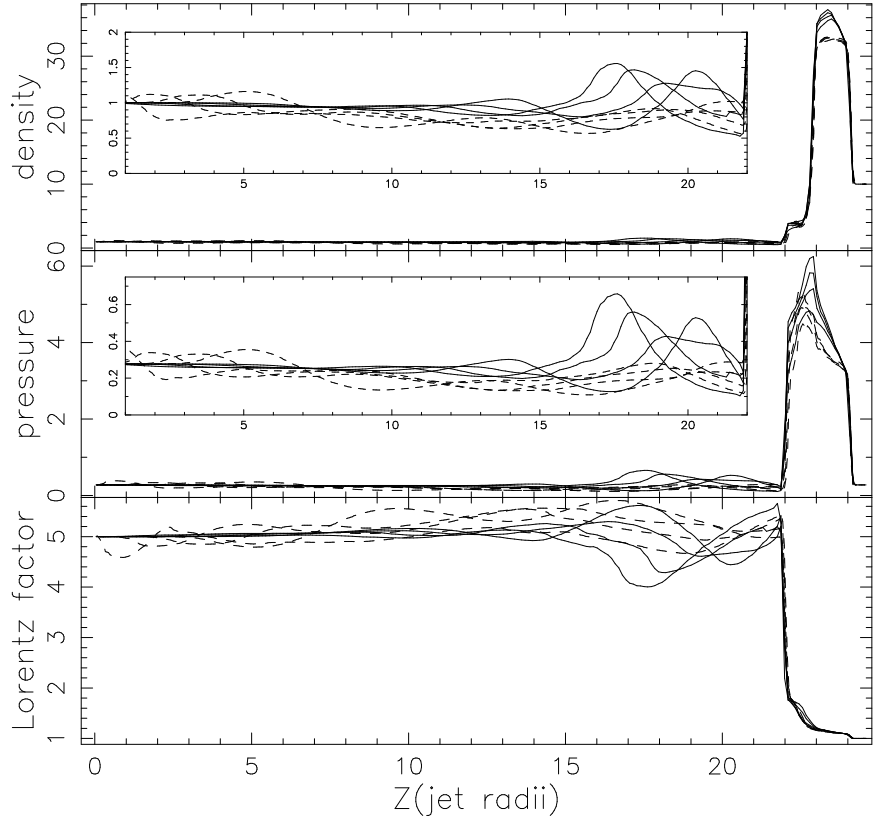}
\caption{
Rest frame density, pressure and Lorentz factor as a function of distance
along the jet axis ($Z$), at $0.2$, $0.4$, $0.6$ and $0.8$ jet-radii, for the
2D simulation (solid lines), and a 3D simulation (dashed lines) of the same
resolution. Insets in the density and pressure panels show the respective
variable upstream of the flow's head. Density and pressure variations
upstream of the Mach disk are of higher amplitude in the 2D case, with peaks
at smaller $Z$ for cuts further from the jet axis, while the 3D case shows a
more widely distributed, but lower amplitude series of variations.
\label{axiquant}}
\end{figure}

In global terms the 2D and 3D simulations are similar: the inclination of the
leading bow shock, a well-defined Mach disk standing back from the contact
surface, and a cocoon that (near to the head) spans at most $5$ jet-radii.
This similarity is quantified by comparing the variation in density, pressure
and Lorentz factor for cuts along the flow axis. Figure~\ref{axiquant}
shows that the global morphology differs little between the 2D and 3D cases.
The only significant difference is seen in the off-axis cuts just upstream of
the head, and reflects the more extensive, but lower amplitude, pressure
wave-driven flow structure in the 3D case as discussed above. This has
minimal influence on the global morphology because the internal structures
are associated with only slight dissipation of the flow energy: the global
dynamics is determined primarily by the energy and momentum flux along the
jet and that is captured as well in 3D as in 2D. The similarity is even more
striking in the dynamics: the average speed of advance of the bow shock is
$0.648$ in the 2D case, and $0.651$ in the 3D case.

\placefigure{axiquant}

These results have an important implication. Our 3D simulations encompassing
the global dynamics (jet, bow and contact) will not be able to achieve a
spatial resolution necessary to capture the flow contact surface well-enough
to reveal it's instability, with the consequent driving of certain normal
modes of the jet. However, in as much as it is the global dynamics of the
source that is of interest, even the modest resolution 3D simulations that
are currently viable do provide a valid picture. We are thus in a position to
address the influence on global morphology of jet deflection through an
encounter with ambient inhomogeneities, or large amplitude precession of the
inflow.  Note that as regards exploring the details of internal jet
structure, we may study a {\it pre-existing} flow, established across the
computational domain in pressure balance with a {\it low density} ambient
medium, representing the cocoon established after the passage of a bow shock.
Within available resources we have performed such simulations with $\sim 25$
cells across the jet diameter, confirmed that almost pure normal modes may be
excited through precession-induced driving at frequencies suggested by linear
stability theory, and demonstrated that linear stability theory correctly
predicts jet structures even for highly nonlinear development ($\delta q \sim
q$) \citep{hhr}. 

\section{Results}

\subsection{Deflection by an Ambient Density Gradient}

 As noted in \S 1, oblique shocks are likely to play a major role in the
understanding of evolving parsec-scale jets, and in the detailed morphology
of kiloparsec-scale flows such as M~87. While models based on transverse
propagating shocks have been extremely successful in explaining both the
radio polarization and broadband behavior of BL~Lac objects and QSOs
\citep{ha1,ha2,mag}, it has become evident recently that only oblique
structures are capable of providing a quantitative explanation for many
features seen on a range of length scales \citep{bab,all}. Oblique shocks
have been widely studied in the context of recollimated flows, e.g.,
\citep{lea,dam} and for their role in mediating bends in supersonic flows,
e.g., \citep{ick,al1,ml1,ml2}.  Such structures are stationary, but
disturbances to the flow are likely to generate propagating oblique
structures also. Here we explore both stationary and propagating oblique
structures that arise due to a change in the jet direction.

 Curvature may result from the nonlinear development of flow instability, as
may internal structure, where, ultimately, modes driven by perturbations to
the jet steepen to form shocks. We shall address this issue in future studies
of the development of instability. Here we are concerned with flow curvature
induced by an external influence, and the associated internal structure that
mediates the change in flow direction.  A jet flow may be bent by a
cross-wind, an ambient pressure gradient, an ambient density gradient, or
interaction with a discrete ambient inhomogeneity (cloud). A cross-wind will
cause a large scale curvature such as seen in NAT and WAT sources
\citep{mpg}, and is unlikely to be the origin of subparsec-scale curvature,
particularly given the implausibility of sustaining a cross wind deep within
a galactic potential. An ambient pressure gradient will also cause large
scale curvature and will relax unless sustained by a gravitational
potential.  In the latter case, the gradient is likely to be in the same
sense as that in which the jet propagates from the central engine, leading to
little or no curvature. We have thus opted  to study the interaction of a
relativistic flow with an ambient density gradient, as this seems the
external influence most likely to cause flow curvature, and which can be
regarded as an idealization of the interaction with a discrete cloud.

An issue with this type of study, known from the simulation of
nonrelativistic jet interaction with clouds \citep{hod}, is that a nontrivial
interaction occurs only for a narrow range of jet parameters. On either side
of this range the momentum flux of the jet is either large enough that the
cloud provides negligible hindrance to the jet flow, or so low that the jet
is effectively stopped by a rigid obstacle. As the thrust of a relativistic
jet increases rapidly with Lorentz factor, we explore in detail only flows
with modest Lorentz factor:  $\gamma<5$.  The ambient density gradient has
been modeled as a smooth ramp running from low density $n_{\rm low}$ to high
density $n_{\rm high}$ away from the inflow, at angle $\phi$ to the inflow
plane, and with ramp width $\Delta L$ measured in units of the jet inflow
radius. The ambient medium is assumed to have uniform pressure distribution,
implying a lower temperature in the high density region; that is consistent
with the high density region modeling a condensation that has achieved
pressure balance with the surrounding material after experiencing thermal
instability. Indeed, observations of the ISM, e.g., \citet{mye}, show that
within each phase, the dispersion in density variations is of order the mean
density, and the range of densities encountered within any phase spans
typically less than an order of magnitude, with even smaller pressure
variations. To assess the maximum likely consequence of such density
variations, we consider a jump in density by one order of magnitude, over a
scale of order the jet radius.

As anticipated, for ambient gradients more-or-less parallel to the inflow
plane the jet material is either deflected to form a bubble of hot gas
encompassed by the slightly distorted interface to the high density domain,
or rapidly penetrates that interface.  The relativistic nature of the flow
does not lead to the formation of distinctive structures. Only interfaces
that are highly inclined to a flow allow the flow to retain its integrity,
and lead to the formation of distinct internal structures. In the following
subsections we present results for a range of flow speeds and density
gradient orientations. We have limited these studies to $\Delta L\sim 1$, as
the jet will respond only weakly to more gradual gradients.

\subsubsection{$\gamma=2.5$, $\phi= 65\arcdeg$}

Figure~\ref{defl25A} illustrates the case $n_{\rm low}=3$, $n_{\rm high}=30$,
$\phi= 65\arcdeg$ and $\Delta L=1.2$ for a jet with $n_{\rm jet}=1$ and
$\gamma=2.5$.  Note that a region of extremely low density and pressure and
high shear (a change in $v_z$ by $0.94c$ over a scale $1.8$ jet-radii)
develops on the low density side of the jet -- immediately to the left of the
inflowing jet in the left panel.  The jet exhibits a number of oblique
structures, primarily driven by the impact of the cocoon, the dynamics of
which are clearly seen in the third panel: the cocoon is one-sided, and
constitutes a structure orthogonal to the jet flow, with marginally
relativistic motion partially directed towards the base of the jet. A large
fraction of the low density cocoon material is moving orthogonal to, and away
from, the jet with a speed $v_x\sim -0.396$, fast enough to exhibit (if seen
face on) modest but significant Doppler boosting (by $\sim 2.25$). That part
of the cocoon that impacts the `base' of the jet does so with a speed
$v_x\sim 0.34$, but the low density and small volume of this portion of the
cocoon mean that the influence of this flow is only to slightly perturb the
jet.  Further evolution would have the jet continue to propagate within the
higher density material, and the head would no longer be aware of the ambient
density gradient. However, until the head is many more jet-radii further on,
the shocked jet material will preferentially follow the channel that takes it
back on one side of the jet, maintaining a source of driving for structures
internal to the jet.

The fourth panel in Figure~\ref{defl25A} shows a grey scale version of the
first panel, colored blue, added to a reddened version of a similar snapshot
4\% of the final time earlier. Unchanged regions are thus rendered neutral,
while the evolved portions are evident as red or blue (or as `ghost' images
if seen in monochrome), and enable us to set limits on the speed of
propagation of the internal jet structures; these we find to be propagating
at a speed $<12\%$ of the marginally relativistic bow shock -- to a good
approximation they are stationary. These structures mediate a change in jet
direction (by $\sim 3\fdg 5$) midway between inflow and Mach disk, which in
view of the approximate stationarity of the structures must persist as the
bow moves well beyond the density interface. Above the `bend' the velocity
component $v_x$ has an average sense and magnitude ($\sim -0.05c$) consistent
with the entire body of the jet following the deviation in flow direction,
while several jet-radii upstream of the Mach disk an oblique structure may be
seen in the left-most panel of Figure~\ref{defl25A} (crossing the jet from
lower left to upper right) which is a shock in which the pressure jumps by
$\sim 2.8$ and which mediates a change in the flow direction so that close to
Mach disk/bow shock the flow has resumed its original direction of
propagation. 

By rendering the laboratory frame density gradient in slices parallel to the
inflow plane, Figure~\ref{defl25B} shows the extent of the cocoon, also
delimited by the twin-peaks in the pressure and density plots of
Figure~\ref{defl25C} which shows plots of Lorentz factor, pressure, rest
frame density and momentum flux along a stack of cuts, such as the one shown
by the horizontal white line in the lower-left panel of
Figure~\ref{defl25B}.  The upper-left panel showing the Lorentz factor makes
clear the slowing of the jet just upstream of the Mach-disk, but also that
the higher pressure there leads to a higher momentum flux -- as seen in the
lower right panel. In fact, the momentum flux increases with $z$, up to and
inclusive of the 11th slice, and manifests significant variation across the
cross-section of the flow at different $z$, as a consequence of the internal
structure. Along a locus of peak Lorentz factor or momentum flux spanning
planes $z=$constant (the result is insensitive to the choice of variable, and
the locus closely follows the inflow axis) a) the Lorentz factor retains its
inflow value ($2.5$) for $12$ jet radii, slowly dipping to $\sim 2$ upstream
of the bow shock at $20$ jet radii; b) the momentum flux varies slowly, by
less than $50$\%, along the first $20$ jet radii, with a value within $10$\%
of its inflow value just upstream of the bow. Interaction with the ambient
medium has lead to the development of discernible internal structure not seen
in the axisymmetric case, but the global dynamics of the jet (if not the
cocoon!) are minimally influenced by this interaction.  We conclude that
inhomogeneity in the ambient medium may lead to significant, oblique
internal jet structure -- albeit essentially stationary -- with little flow
energy dissipation, or change in the overall direction of the flow.
Stationary VLBI structures, which may be a manifestation of such oblique
shocks, are now known to be very common \citep{jor}.

\placefigure{defl25A}

\placefigure{defl25B}

\placefigure{defl25C}

\subsubsection{$\gamma=5.0$, $\phi= 65\arcdeg$}

Figure~\ref{defl50A} shows the evolution of a faster jet ($\gamma=5$) with
the same inclined density gradient. The behavior is not dramatically
different from that exhibited by the $\gamma=2.5$ flow, but a number of
important features may be seen more clearly here. As the flow meets the
density gradient, a clockwise rotation of the Mach disk occurs (upper panels
in Figure~\ref{defl50A}), while after the initial interaction, the Mach disk
rotates counter-clockwise (bottom panels in Figure~\ref{defl50A}) to lie
orthogonal to the density incline by the last cycle of computation.  The
maximum  clockwise rotation of the disk is $\sim 12\arcdeg$, and the maximum
counter-clockwise rotation is $\sim 21\arcdeg$: highly polarized emission
from the downstream flow of the Mach disk may well be the dominant feature on
VLB maps, but remain unresolved, the direction of the electric vector being
used to establish the likely flow direction; deviations from the transverse,
by tens of degrees, could lead to a significant misestimation of the local
flow direction. The jet deflection is mediated by an internal shock whose
development starts to become evident in the fifth panel as a structure
crossing the Mach disk.  By the last cycle, this has developed to form an
internal structure extending upstream from the Mach disk, with an inclination
comparable to that of the density gradient. The higher momentum flux of the
jet in this case ensures that deflection occurs only within the extent of
this oblique shock, and the evolution is not followed far enough to see
realignment of the jet along its original direction of propagation. The
internal shock that moderates the jet deflection is quite strong, the rest
frame density and pressure jumps being $\sim 5$ and $\sim 14$ respectively.

\placefigure{defl50A}

\subsubsection{$\gamma=10.0$, $\phi= 65\arcdeg$}

Figure~\ref{defl10A} shows the evolution of a $\gamma=10$ jet with the
same inclined density gradient. The speed and jet-to-ambient density contrast
are comparable to those of cases C and D studied by \citet{dah} wherein it
was found that the ambient medium presented little obstruction to the flow,
and little internal jet structure arose. In the present case the jet head
moves with high efficiency ($\sim 0.95$; see \citet{ros}) through even the
densest ambient gas at relativistic speed, and all sections of the head cross
the density incline before a significant rotation of the Mach disk can occur.
A rotation of the Mach disk, deflection of the jet, and generation of oblique
internal shocks occurs only if the jet is of low enough speed that the
densest ambient regions lead to an inefficient, slow head advance there.

\placefigure{defl10A}

\subsubsection{$\gamma=5.0$, $\phi= 35\arcdeg$}

Figure~\ref{defl50sA} shows the evolution of a $\gamma=5$ jet with a
density gradient inclined at a shallower angle -- $35\arcdeg$ -- than
explored previously. As would be anticipated the clockwise/counter-clockwise
rotation of the Mach disk as the jet crosses the interface is slight.
However, this shallower angle case sheds light on the origin of the jet
deflection. A conical pressure wave which steepens to form a weak conical
shock is driven {\it asymmetrically} into the jet flow at the location of the
ambient gradient, and the asymmetric nature of the flow perturbation leads to
its deflection. Some jet-radii beyond this point an oblique internal shock
that cuts the Mach disk returns the flow to its original direction. The
inclination of the density gradient allows us to follow the jet further into
the denser ambient medium in this case. By the end of the simulation both the
morphology of the cocoon and the velocity field within it show that the
shocked jet flow is similar -- albeit still somewhat asymmetric -- to that
seen for propagation in a uniform environment. However, without the low density
region to act as a sink for the backflowing shocked jet material, the cocoon
at left, just behind the head, starts to inflate. That induces a curvature to
the contact surface between shocked jet and shocked ambient material,
enhanced by the Kelvin-Helmholtz instability, and this in turn drives a
pressure wave into the jet just upstream of the Mach disk. A normal mode of
the jet is excited (cf. \citet{har}) which steepens to form a conical shock
upstream of the Mach disk. The influence on the flow is quite significant:
constriction of the flow accelerates the jet upstream of the conical shock to
$\gamma\sim 6$, but the shock decelerates the flow to $\gamma\sim 4$; the
corresponding pressure jump is $\sim 8.5$.

\placefigure{defl50sA}

In as much as the types of interaction just discussed are uniquely
associated with the leading edge of jets, the structures that result are
localized, and will not form a web of propagating shocks along the length of
a flow. However, this study highlights that the Mach disk associated with
the hotspots of kiloparsec (and larger) scale jets may display orientations
-- revealed by their polarization structure -- that bears no simple relation
to the jet orientation and accompanied by emission from associated oblique
shocks, if the head experiences a significantly inhomogeneous ambient
medium.  Further, on any scale, and in particular on the unresolved
sub-parsec scale, a jet that restarts after a period of quiescence that
allows a previously formed channel to relax, or that restarts with a new
orientation, will display polarized emission that is a complex sum of that
from the rotating Mach disk and developing internal shocks, and which on such
scales will evolve within observable time frames.

\subsection{Precession}

 Symmetries seen in twin-jet sources, in particular the S-symmetric sources,
e.g., \citet{lea}, and the intriguing success of binary black hole models for
the optical-radio waveband behavior of OJ~287, e.g., \citet{val}, suggest
that jets are subject to a precessional motion, and provide a quantitative
explanation for the possible origin of that motion, respectively. Even if
precessional motion associated with a massive binary system is not the cause
of a time-dependent curvature seen in some parsec-scale flows (e.g., BL~Lac;
Mutel et al., private communication), an exploration of the impact of
precession on flow structure and evolution will give insight into the
internal structures to be found in evolving, curved flows.

 As a first step in exploring the role of precession, we have applied
precession to the inflow of the axisymmetric jet discussed above: a jet with
inflow Lorentz factor of $5$.  The inflow precesses on a cone of semi-angle
$11\fdg 25$ with a frequency $0.2885$ rad measured in time units set by the
inflow radius and speed. The relatively large computational domain, $\sim
32\times 32\times 82$ jet-radii, meant that we could employ only three grid
levels, with refinement by $\times 3$, providing $\sim 11$ fine cells across
the inflowing jet diameter.  In the approximation that both the inflow speed
and bow propagation speed are close to unity, the precession rate implies
$\sim 3.75$ revolutions of the jet during evolution across the computational
domain. The computation was terminated when the bow was $\sim 90$\% of the
way to the domain's far edge, as by then significant dissipation and
disruption of the jet had occurred. Renders in planes parallel to the inflow
plane reveal a small influence of the boundary on structures close to the
edge of the computational domain. However, there is no evidence that the
limited lateral extent of the computational domain influences the spine of
the flow, which is that structure of most significance to us.

 Figure~\ref{precA} shows a sequence of slices in the plane $x=0$ at
equally-spaced intervals during the evolution, using schlieren plots of
laboratory frame density. Superficial inspection of the final slice suggests
that to a significant extent, the jet has retained its integrity, and is
driving a bow not dissimilar to that seen in the unprecessed case. In
particular, on average the bow is advancing at $v\sim 0.41$, which is $\sim
63$\% the speed with which it advanced in the unprecessed case, suggesting a
relatively undiminished jet thrust. However, Figure~\ref{precB} shows
from left to right, a schlieren render of the pressure -- there are dramatic
variations of pressure along the jet's path, suggestive of significant
dissipation; a linear render of the pressure (dominated by the leading bow)
with 3-velocity vectors superposed -- evidently the jet's momentum has been
shared with a broad sheath of material; and finally, the Lorentz factor --
showing that despite precession, the jet does retain its integrity for almost
$50$ jet-radii, but thereafter the flow speed drops dramatically. The
asymmetric bow is being driven forward by a flow in which the pressure
maintains a high enthalpy, and which is to some extent focussed onto a small
area at the bow's apex.

Figure~\ref{precC} quantifies this, showing the run of rest frame
density, pressure, Lorentz factor and momentum flux (or `discharge'), ${\cal
F}=\gamma^2\left(e+p\right) v_z^2+p$, along the spine of the flow.  This
spine was defined by computing weighted, average $x$ and $y$ values in a
series of planes $z=$constant, within a cylinder of diameter $6$ jet-radii
aligned with the jet inflow (to exclude complex structures near to the edge
of the computational domain). Using the Lorentz factor as a weighting function
produced a locus barely distinguishable from that found using the momentum
flux as a weighting function, leading us to the conclusion that this approach
indeed picks out a physically significant core flow. The quantities plotted in
Figure~\ref{precC} are spatial averages within cylindrical pills of one
jet radius, aligned locally with the spine. Increasing the radius of these
sampling volumes decreased the average Lorentz factor and momentum flux, but
did not change the character of the variation of these quantities along the
spine. Beyond $50$ jet-radii the Lorentz factor falls well below its initial
value and the local flow direction becomes more chaotic.

Indeed, the orientation of the velocity vectors within the jet, relative to
the local jet `direction' -- defined by the locus of the spine -- is a
quantity of fundamental importance. Is the flow locally parallel to the jet
boundary, with little momentum imparted to the jet's wall, leading to the
expectation that flow disturbances (e.g., shocks) would follow the jet
channel, or do fluid elements move more ballistically? The bottom panel of
Figure~\ref{precC} shows the angle between the spine of the flow and the
inflow direction (i.e., the $z$-direction) as a dashed line, and the angle
between the spine of the flow and the velocity as a solid line. The former
changes quite smoothly along the flow, increasing to a value well above that
imposed at inflow, in conjunction with a decline in the $z$ component of
momentum flux. By $\sim 50$ jet-radii large fluctuations associated with jet
disruption become evident. If the velocity vectors tracked the flow, the
solid line would remain close to zero. However, until significant disruption
of the jet is evident, the angle between the spine of the flow and the
velocity closely follows, but lies a little below, the angle of the spine.
This is because the velocity vectors are almost entirely along the $z$-axis,
with a slight offset (typically $\la 4\arcdeg$) towards the local channel
direction (with the potential to change the Doppler boost of emitted
radiation by more than $25$\%).  Thus while there is a weak tendency for
velocities to follow the jet channel, the motion is substantially ballistic,
with significant momentum imparted to the jet wall, displacing that surface
as the flow evolves.  Figure~\ref{precD} shows the evolution of the
location of the spine in 9 equally separated planes $z$=constant to a little
beyond $50$ jet-radii.  (The jet spends too little time further from the
inflow to study its evolution there.) The complex pattern of dashed lines
that join these points arises because the spine is captured at various phases
of precession in the 28 epochs sampled, but the stable, conical path of the
spine is evident for tens of jet radii from the inflow.

\placefigure{precA}

\placefigure{precB}

\placefigure{precC}

\placefigure{precD}

Interaction between the precessing flow and its cocoon of shocked jet
material has imparted a significant `forward' velocity to the latter. Within
the broad region delimited by the contact surface between shocked jet and
shocked ambient material the entire flow moves forward at a marginally
relativistic speed, forming a plateau about a spine of highly relativistic
flow.  The corresponding velocity distribution for the unprecessed case shows
significant forward flowing cocoon material only within a jet's radius of the
jet itself. This contrasts even more dramatically with the results seen in
highly underdense jet simulations, e.g., \citet{kf1}, where the cocoon
material shows significant {\it backflow} in the laboratory frame (albeit
away from the jet, near to the contact surface), with the potential for
significant Doppler enfeeblement of its radiation flux. The broad forward
flow seen in the current simulation is suggestive of the core-sheath
morphology suggested to explain observations implying both relativistic and
nonrelativistic motions in the same source, e.g., \citet{has}, and the recent
observation of a source with a `spine-sheath' morphology, evident in the
polarization map \citep{arw}.  Whether jets that propagate from the
sub-parsec scale to the tens of kiloparsec scale can retain their integrity
under the influence of large amplitude precession can be answered only
through larger scale computations involving a diverging flow, wherein the
local scale length to jet radius ratio is bounded.

\section{Discussion}

\begin{figure}
\epsscale{1.0}
\plotone{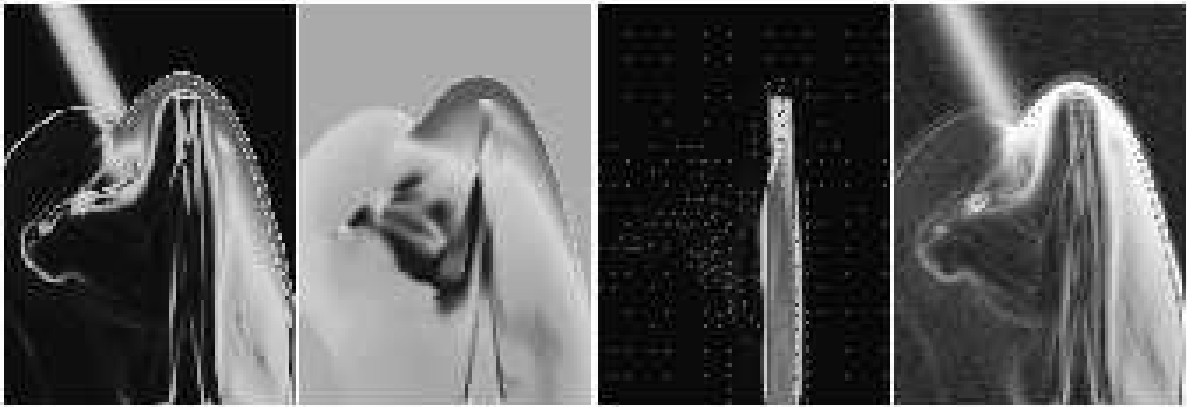}
\caption{
A slice orthogonal to the inflow plane, showing a $\gamma=2.5$ jet
interacting with an oblique ($65\arcdeg$) ambient density gradient in which
the density increases to the right by $\times 10$ across a region of scale
length $1.2$ jet-radii.  From left to right: the gradient of the laboratory
frame density; the pressure; the Lorentz factor with 3-velocity vectors
superposed; superposed, color-coded renditions of the left-most panel at two
late epochs, showing the motion of flow features: in a monochrome render the
motion of the bow and Mach disk show up as a `ghost' images. The peak Lorentz
factor of $2.5$ is rendered red, and values of order $1$ are dark blue -- the
peak values appear as the interior gray `stem' in a monochrome render.
\label{defl25A}}
\end{figure}

\begin{figure}
\plotone{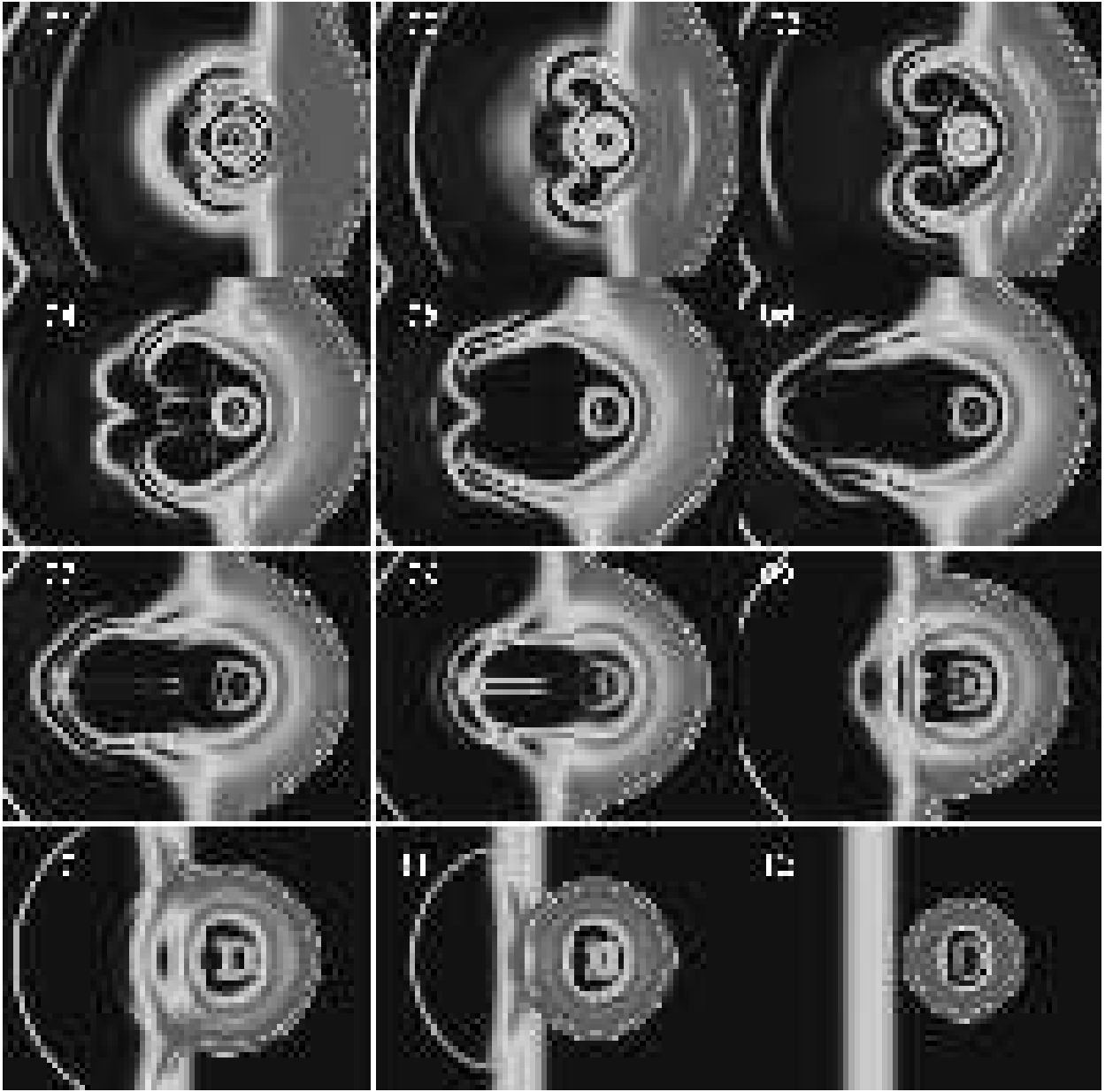}
\caption{ As for Figure~\ref{defl25A}, showing the gradient of the
laboratory frame density in planes parallel to the inflow surface, at 12
equally spaced locations along the $z$-axis between the inflow plane and the
Mach-disk that terminates the jet.
\label{defl25B}}
\end{figure}

\begin{figure}
\plotone{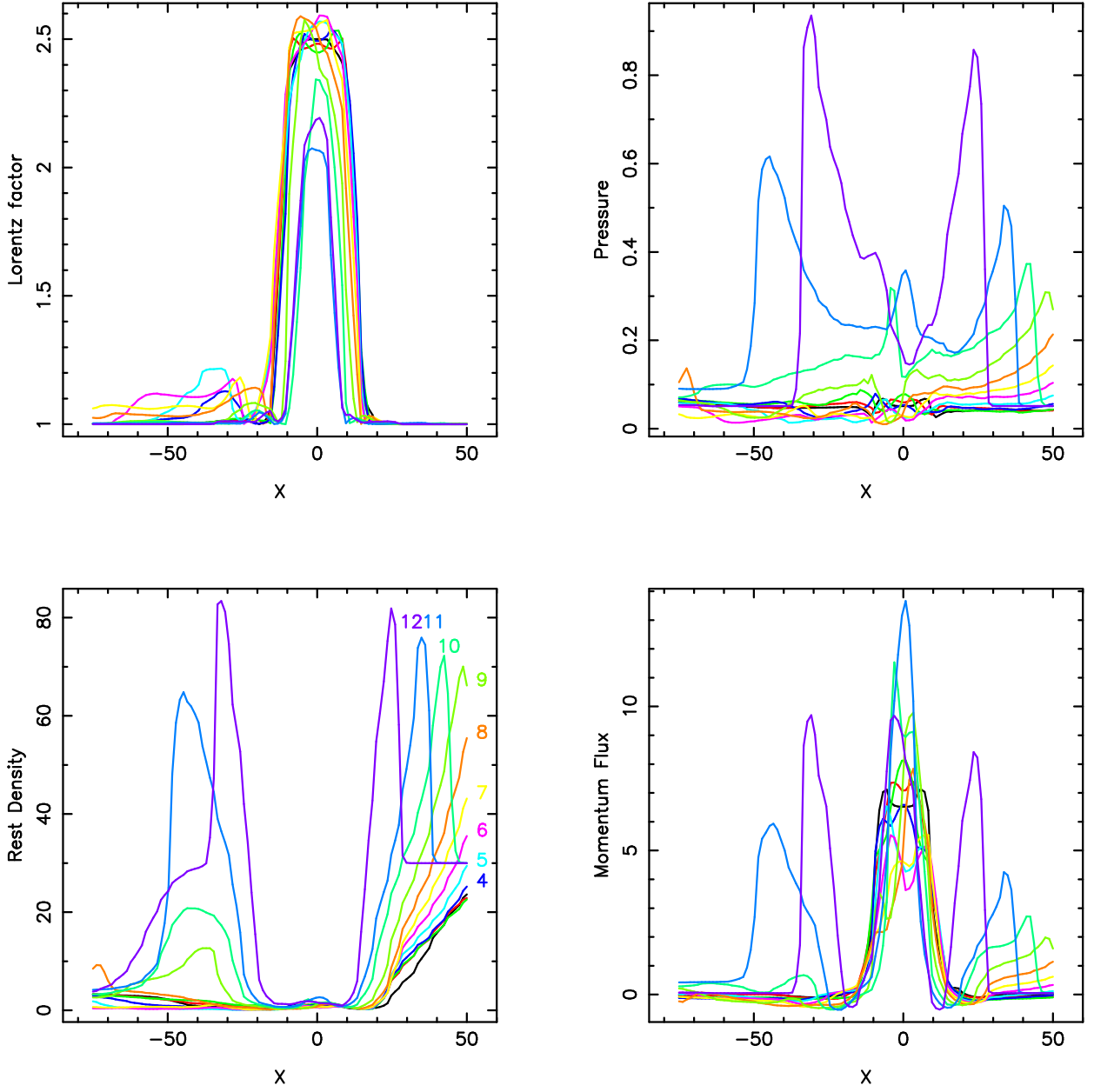}
\caption{As for Figure~\ref{defl25A}, showing the Lorentz factor,
pressure, rest frame density and momentum flux (${\cal
F}=\gamma^2\left(e+p\right) v_z^2+p$) along the cut shown as a white line in
the lower-left panel of Figure~\ref{defl25B}, at 12 equally spaced
locations along the $z$-axis, between the inflow plane and the Mach-disk that
terminates the jet. The numeric labels in the lower-left panel correspond to
the sequence of $z$-values shown in Figure~\ref{defl25B}.
\label{defl25C}}
\end{figure}

\begin{figure}
\plotone{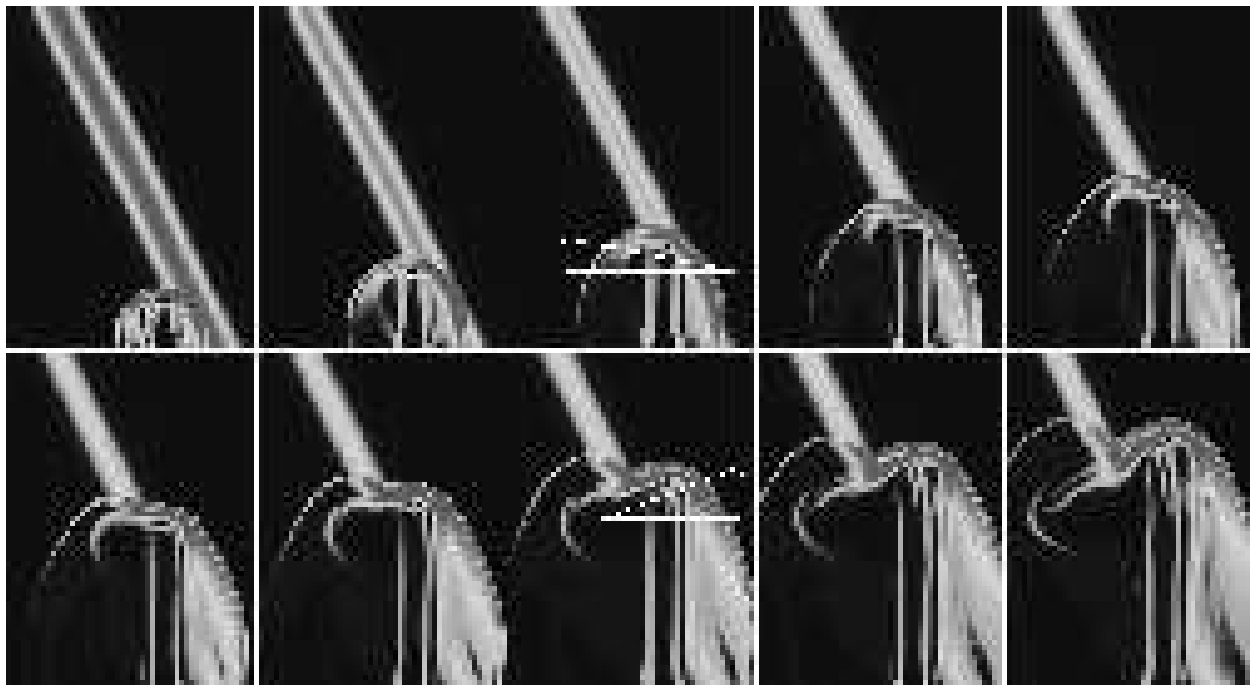}
\caption{Slices orthogonal to the inflow plane, showing a $\gamma=5.0$ jet
interacting with an oblique ($65\arcdeg$) ambient density gradient in which
the density increases by $\times 10$ across a region of scale length $1.2$
jet-radii. The gradient of the laboratory frame density is rendered.
Lines in the third and eighth panels indicate the sense and magnitude of
the rotation of the Mach disk.
\label{defl50A}}
\end{figure}

\begin{figure}
\plotone{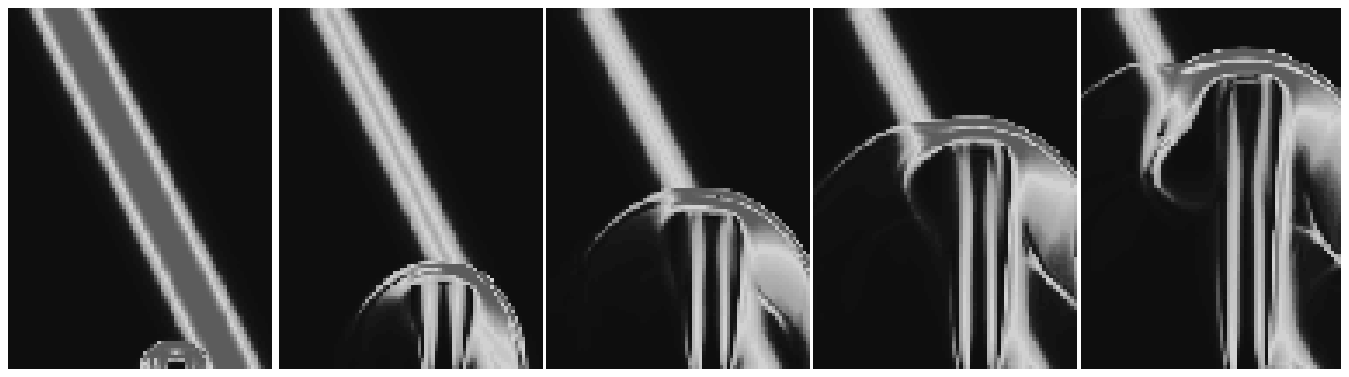}
\caption{Slices orthogonal to the inflow plane, showing a $\gamma=10.0$ jet
interacting with an oblique ($65\arcdeg$) ambient density gradient in which
the density increases by $\times 10$ across a region of scale length $1.2$
jet-radii. The gradient of the laboratory frame density is rendered.
\label{defl10A}}
\end{figure}

\begin{figure}
\plotone{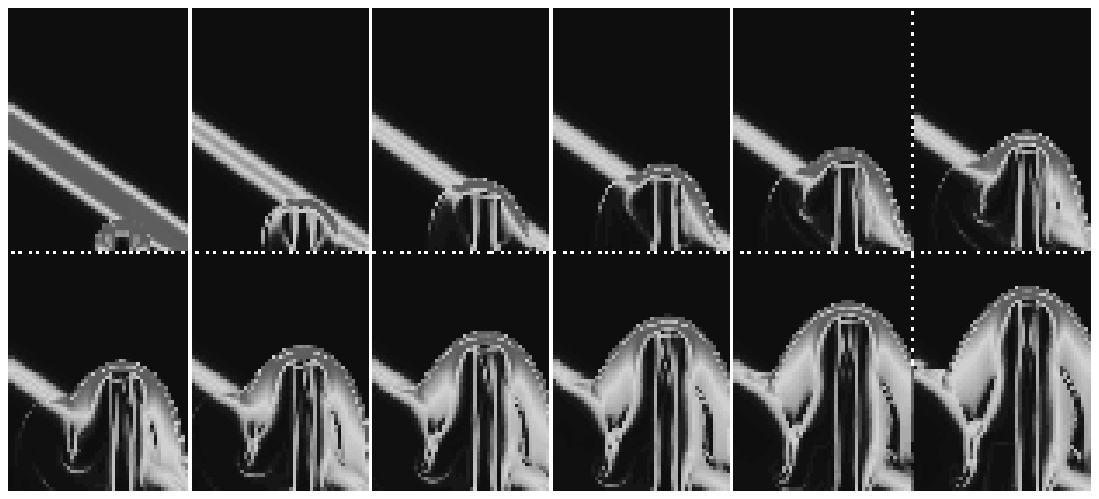}
\caption{Slices orthogonal to the inflow plane, showing a $\gamma=5.0$ jet
interacting with an oblique ($35\arcdeg$) ambient density gradient in which
the density increases by $\times 10$ across a region of scale length $1.2$
jet-radii. The gradient of the laboratory frame density is rendered.
\label{defl50sA}}
\end{figure}

 We have shown that jets impacting an inclined ambient density gradient are
subject to the development of internal waves and oblique shocks. It is
impractical to run the simulations to a point where the leading bow shock
leaves the computational domain, but we have shown that these internal
structures are slow moving, and argued that their driving flows persist long
after the bow has passed the deflecting structure; thus an interesting
question is the extent to which these internal structures would be observable
in maps not dominated by the bow shock (which even if in proximity to the
structures of interest, might contribute little to the synchrotron emissivity
-- e.g., \citet{mhd}). In the simulation with $\gamma=2.5$, $\phi= 65\arcdeg$
(see Figure~\ref{defl25A}) the oblique, high pressure structure upstream of
the flow's head spans the jet in the plane parallel to the ambient density
gradient, and in a plane orthogonal to that, constitutes a wedge of extent
$\sim 2.75$ jet radii, and maximum width $\sim 0.43$ jet radii. It thus fills
$\sim 7$\% of the volume of the perturbed jet segment. The peak pressure in
this structure is $p_{peak}\sim 0.38$, while in the adjacent cocoon
$p_{ext}\sim 0.2$, and in the initial state $p_{int}=0.05$. If the emission
is optically thin, with spectrum $\propto \nu^{-\alpha}$, there is
equipartition between radiating particles and random magnetic field ($u_e\sim
u_B$), and the number of radiating particles is proportional to the internal
energy and thus the pressure, the emissivity is $j\propto
p^{\left(\alpha+3\right)/2}$ \citep{mhd}. For $\alpha=0.5$ the emissivity of
the oblique structure is $\sim 35$ times that of the quiescent jet.  Allowing
for the $\sim 7$\% filling factor, this structure would be $\sim 2.4$ times
brighter than a corresponding segment of unperturbed jet. The simulation with
$\gamma=5$ and $\phi= 35\arcdeg$ (see Figure~\ref{defl50sA}) also shows a
prominent oblique internal structure in the final time frame; again, the
structure spans the jet in a plane parallel to the ambient gradient and its
section orthogonal to this direction approximates an ellipse with major axis
tilted slightly with respect to the jet's inflow direction. The filling
factor is $\sim 50$\%. The corresponding pressure values are $p_{peak}\sim
0.78$, $p_{ext}\sim 0.2$ and $p_{int}=0.27$ (the initial pressure being
higher than in the slow jet case for fixed Mach number). Following the same
arguments as used above to estimate the component flux, it follows that this
structure would be $\sim 3.2$ times brighter than a corresponding segment of
unperturbed jet. We conclude that where jets are disturbed by ambient density
structure, jet emission may be enhanced by a significant factor ($2-3$)
locally, with little impact on the jet's global dynamics.

\begin{figure}
\epsscale{0.4}
\plotone{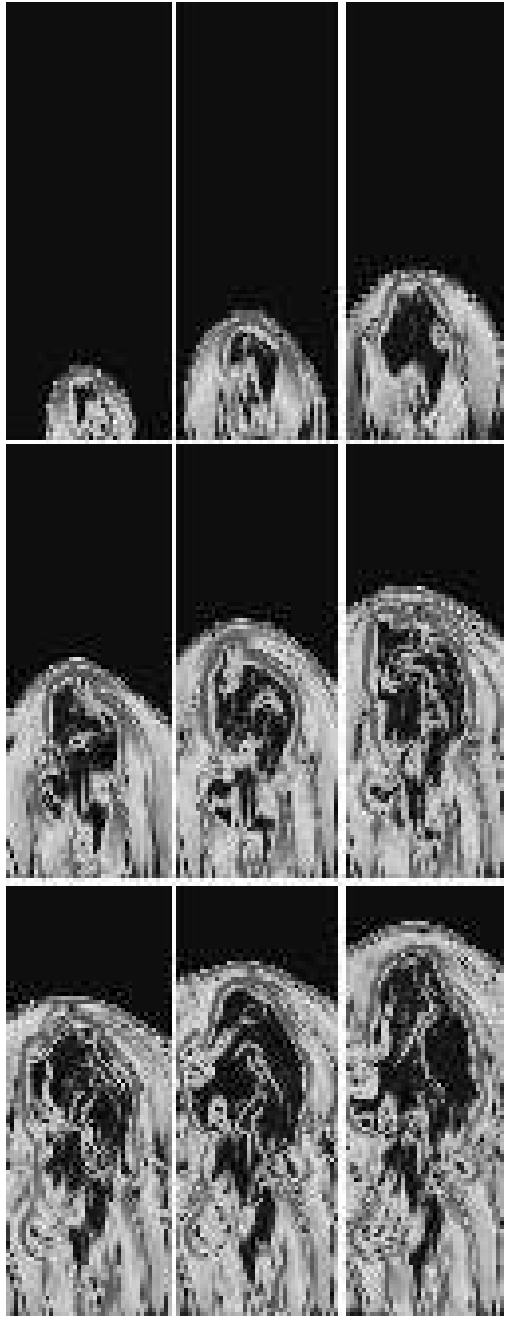}
\caption{
Slices in the plane $x=0$ at equally-spaced intervals during the evolution of
the precessed $\gamma=5$ jet. Each panel is a schlieren rendering of the
laboratory frame density.
\label{precA}}
\end{figure}

\begin{figure}
\epsscale{1.0}
\plotone{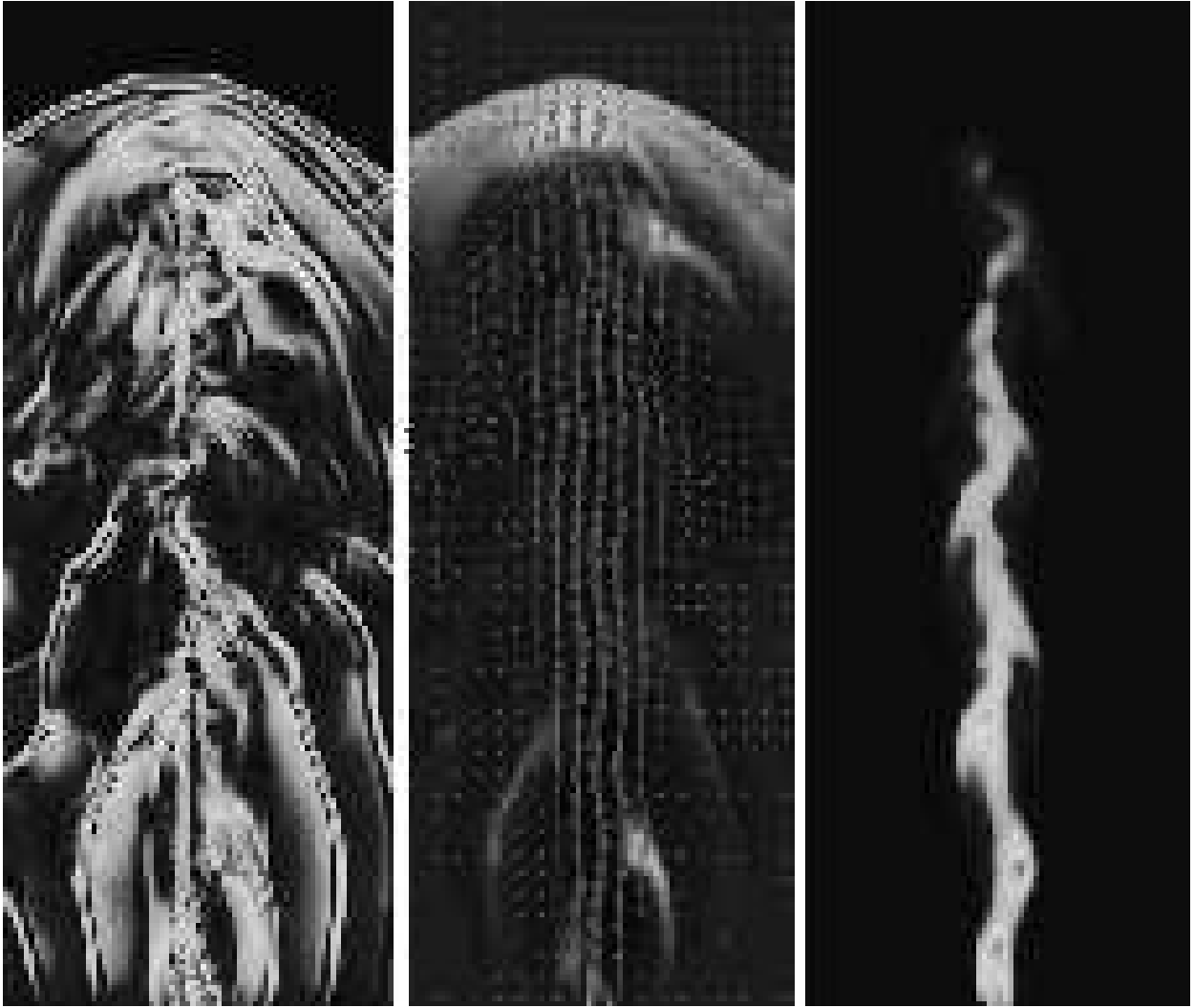}
\caption{
Slices in the plane $x=0$ at the last time step for the precessed $\gamma=5$
jet, showing from left to right: a schlieren render of the pressure, a linear
render of the pressure with 3-velocity vectors superposed, and the Lorentz
factor; the peak Lorentz factor of $5$ is rendered red, and values of order
$1$ are dark blue -- the peak values appear as dark regions in a monochrome
render. The jet retains its integrity for $\sim 50$ jet-radii, and thereafter
rapidly disrupts.
\label{precB}}
\end{figure}

\begin{figure}
\plotone{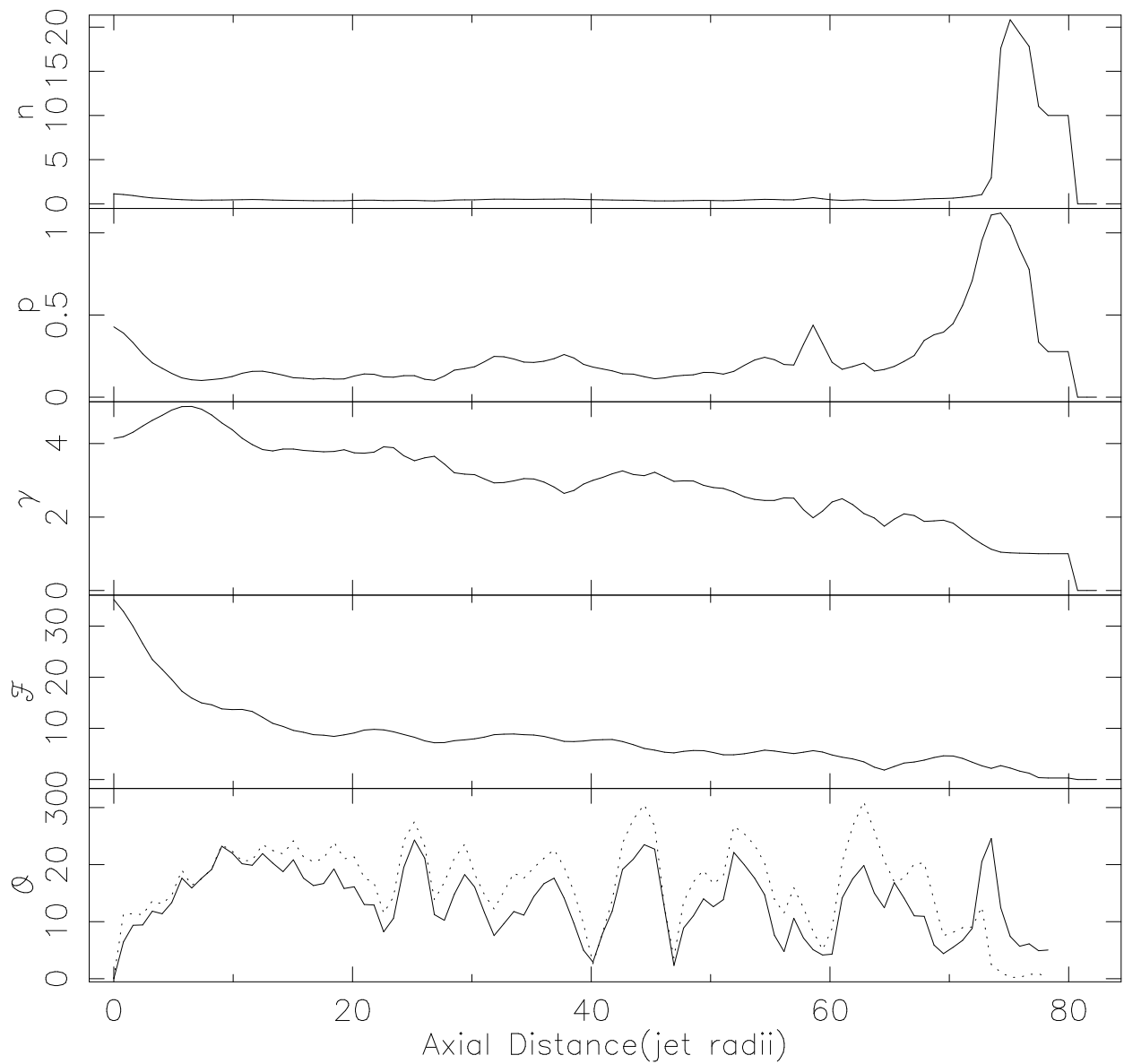}
\caption{
From top to bottom: the run of rest frame density ($n$), pressure ($p$),
Lorentz factor ($\gamma$) and momentum flux ($\cal F$)  along the spine of
the precessed flow. The bottom panel shows the local angle between the spine
and the inflow direction (dashed line) and the local angle between the
velocity vector and the spine (solid line). The plot in the lowest panel is 
terminated where the flow direction becomes ill-defined.
\label{precC}}
\end{figure}

\begin{figure}
\plotone{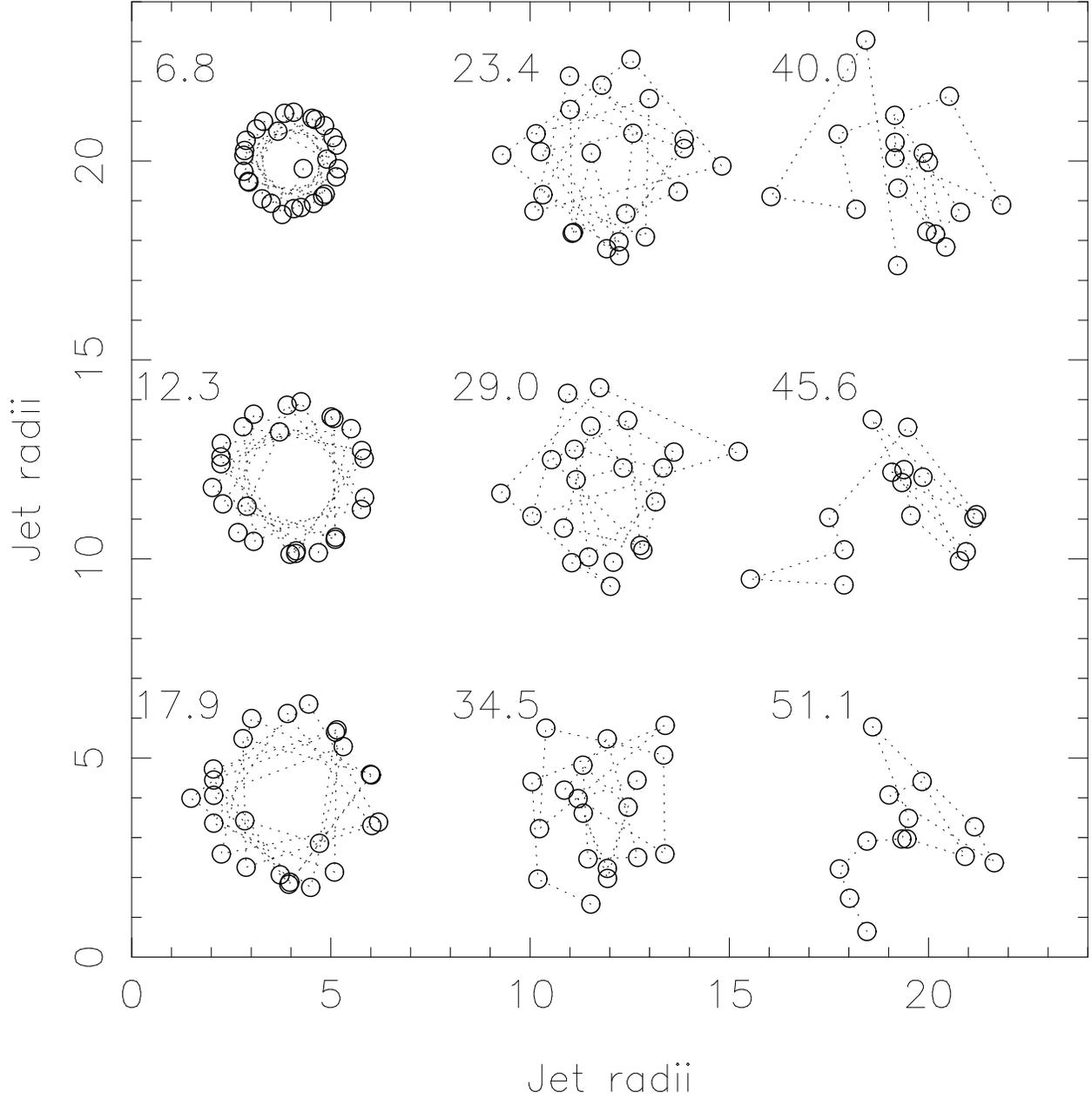}
\caption{
The evolution of the spine of the precessed jet, at 9 locations $z=$constant
labeled in jet-radii. Each mark shows the location of the spine at one of 28
time slices, and the sense of evolution is given by the joining dashed
lines.
\label{precD}}
\end{figure}

 A simple prescription for the appearance of the precessed jet (see
Figure~\ref{precA}) cannot be readily given, as the structure is highly
asymmetric, and while the local jet boundary evolves at subrelativistic
speed, significant Doppler boosting is associated with the instantaneous flow
velocity, and varies with observer location, even for observers with the same
inclination to the inflow axis. Maps are being computed by C. M. Swift,
following the approach described by \citet{mhd} and these results will be
presented elsewhere.  For the purpose of characterizing brightness variations
that might be associated with such flows, here we consider the variation of a
simple estimator of the source intensity along the flow's spine. We have
computed an effective emissivity ${\cal E}=p^{\left(\alpha+3\right)/2}{\cal
D}^2$ as a function of distance along the spine, where ${\cal
D}=1/\gamma\left(1-\beta\cos\theta\right)$, the Doppler factor, and the boost
is appropriate for stationary, bounded flows \citep{tim}. The well-defined,
high Lorentz factor channel evident in Figure~\ref{precB} illustrates that at
least within a jet radius of the spine, the internal flow conditions are
quite uniform, and that the significantly Doppler boosted flow suffers little
divergence before disruption just upstream of the bow. The observed intensity
will be determined by integrating the emissivity across the Doppler boosted
flow for some line of sight, and given the cross-flow uniformity, and
comparable line of sight depth for each flow segment, ${\cal E}$ provides a
reasonable estimator of the source brightness. We have adopted $\alpha=0.5$,
and an angle of view with respect to the inflow axis $\theta'=10\arcdeg$,
the latter being of interest as the critical cone of the inflowing jet (with
$\gamma=5.0$) is $\sin^{-1}\left(1/\gamma\right)\sim 11.5\arcdeg$. The
detailed run of ${\cal E}$ along the spine depends upon the azimuthal angle
chosen.  However, for all cases, a narrow peak in Doppler factor occurs
between $10$ and $20$ jet radii along the spine, and convolved with the broad
pressure peak at that location (see Figure~\ref{precC}) always leads to a
substantial intensity enhancement:  the peak `intensity' is typically $6$ to
$10$ times that in the `intensity' minimum immediately downstream.  Beyond
$20$ jet radii, significant (by a factor of $2$ or more) fluctuations in
`intensity' occur, the details of which depend on observer orientation, but
which are {\it not} uniquely tied to variations in the underlying pressure
distribution or Doppler factor -- although there is a trend for pressure
fluctuations to underpin the `intensity' variations as the head is
approached. 

The flow thus exhibits a core-jet morphology, in which individual features
can move both towards and away from the jet's origin -- according to the
phase of precession -- by a number of jet radii. In particular, if the core
is associated with the inner-most region where emissivity and high Doppler
boost conspire to produce a high intensity component, the location of that
component will not be stable to within a few jet-radii. As the precession
time scale is likely to be long compared with the time scale for propagation
of an individual component along the flow, this will have minimal effect on a
sequence of VLBI measurements that define the trajectory of one component.
However, it could lead to long-term changes in the relative position of
stationary components, and change the inner jet orientation so as to
influence the details of the interaction of propagating and stationary
components, e.g., as seen in $4C~39.25$ \citep{al2}.  This precessed flow has
a number of characteristics exhibited by BL~Lac objects: a core-jet
morphology, with stationary, or slowly moving components, and a flow that
appears to dissipate rather close to the core.  Jets associated with QSOs may
typically be followed further than those associated with BL~Lac objects, so
perhaps one facet of the class distinction relates to the amplitude of
precessional perturbation to which the jet is subjected. Evidently a small
helical perturbation (transverse velocity of order 1\% of longitudinal
velocity) does {\it not} lead to disruption of the flow \citep{hhr}. Our
immediate goal is to generate a well-sampled grid of simulations for a range
of Lorentz factor and precession cone semi-angle, to establish both the
largest amplitude of precession for which flows of a given Lorentz factor
avoid macroscopic break-up, and the physical origin of that break up. The
extent to which propagating shocked fluid, thought to explain moving VLBI
components, will follow the helical channel, or follow approximately
ballistic trajectories, is also a subject for future study.

\acknowledgments

This work was supported in part by NSF grants AST 9617032 and PHY 9979985, by
NRAC allocation MCA93S025 at NCSA, and by the Ohio Supercomputer Center.
We thank the referee for many useful comments that helped to improve the
presentation of this work.

\clearpage

\clearpage

\clearpage
 
\clearpage
 
\clearpage

\clearpage
 
\clearpage
 
\clearpage
 
\clearpage
 
\clearpage
 
\clearpage
 
\clearpage
 
\clearpage
 
\clearpage
 
\clearpage
 
\end{document}